\documentclass[a4paper,11pt]{article}
\pdfoutput=1 

\usepackage{jheppub} 

\usepackage[T1]{fontenc} 

\usepackage{lineno}

\def\Ecm {\ensuremath{\rm E_{\rm c.m.}}}

\def\epem {\ensuremath{e^+ e^-}}
\def\pipi {\ensuremath{\pi^+\pi^-}}

\def\mevcc {\ensuremath{\rm MeV/c^2}}
\def\piz {\ensuremath{\pi^0}}
\def\ppz {\ensuremath{\pi^0\pi^0}}

\title{\bf{ \boldmath
Study of the process $e^+e^-\to \pi^+\pi^-\pi^0\pi^0\eta$
in the energy range 1.6--2.0 GeV with the CMD-3 detector
}}

\author[a,b]{R.R.\,Akhmetshin}
\author[a,b]{A.N.\,Amirkhanov}
\author[a,b]{A.V.\,Anisenkov}
\author[a,b]{V.M.\,Aulchenko}
\author[a]{N.S.\,Bashtovoy}
\author[a]{E.V.\,Bedarev}
\author[a]{D.E.\,Berkaev}
\author[a,b]{A.E.\,Bondar}
\author[a]{A.V.\,Bragin}
\author[a]{V.S.\,Denisov}
\author[a,b]{D.A.\,Epifanov}
\author[a,c]{L.B.\,Epshteyn}
\author[a,b]{A.L.\,Erofeev}
\author[a,b]{G.V.\,Fedotovich}
\author[a,c]{A.O.\,Gorkovenko}
\author[a,b]{A.A.\,Grebenuk}
\author[a,b]{S.S.\,Gribanov}
\author[a,c]{D.N.\,Grigoriev}
\author[a,b]{F.V.\,Ignatov}
\author[a,b]{V.L.\,Ivanov}
\author[a]{A.S.\,Kasaev}
\author[a]{S.V.\,Karpov}
\author[a,b]{V.F.\,Kazanin}
\author[a,b]{I.A.\,Koop}
\author[a,b]{A.A.\,Korobov}
\author[a,c]{A.N.\,Kozyrev}
\author[a,b]{P.P.\,Krokovny}
\author[a,b]{A.S.\,Kuzmin}
\author[a,b]{I.B.\,Logashenko}
\author[a,b]{P.A.\,Lukin}
\author[a,b]{K.Yu.\,Mikhailov}
\author[a,b]{I.V.\,Obraztsov}
\author[a]{Yu.N.\,Pestov}
\author[a,b]{E.A.\,Perevedentsev}
\author[a]{V.G.\,Petrochenko}
\author[d]{N.A.\,Petrov}
\author[a,b]{A.S.\,Popov}
\author[a,b]{Yu.A.\,Rogovsky}
\author[a]{A.A.\,Ruban}
\author[a,b]{A.E.\,Ryzhenenkov}
\author[a,b]{A.V.\,Semenov}
\author[a]{A.I.\,Senchenko}
\author[a,b]{V.E.\,Shebalin}
\author[a]{D.N.\,Shemyakin}
\author[a,b]{B.A.\,Shwartz}
\author[e]{D.B.\,Shwartz}
\author[a,b,1]{E.P.\,Solodov\note{Corresponding author: solodov@inp.nsk.su}}
\author[a]{M.V.\,Timoshenko}
\author[a]{V.M.\,Titov}
\author[a,b]{A.A.\,Talyshev}
\author[a,b]{S.S.\,Tolmachev}
\author[a]{A.I.\,Vorobiov}
\author[a]{I.M.\,Zemlyansky}
\author[a]{D.S.\,Zhadan}
\author[a,b]{Yu.V.\,Yudin}

\affiliation[a]{Budker Institute of Nuclear Physics, SB RAS, 
Novosibirsk, 630090, Russia}
\affiliation[b]{Novosibirsk State University, Novosibirsk, 630090, Russia}
\affiliation[c]{Novosibirsk State Technical University, Novosibirsk, 630092, Russia}
\affiliation[d]{Institute for Nuclear Research, RAS, Moscow, 117312, Russia}
\affiliation[e]{P-cure Ltd, Shilat, 7318800, Israel}

\vspace{0.7cm}
\abstract{
The cross section of the process $e^+e^- \to \pi^+\pi^-\pi^0\pi^0\eta$ has been 
measured using a data sample with the integrated luminosity of 372 pb$^-1$
collected 
with the CMD-3 detector at the VEPP-2000  $e^+e^-$ collider.  6300$\pm$145 
signal events have been selected  in the center-of-mass energy range 
1.6--2.0 GeV. The total systematic uncertainty of the cross section is
about 10\%. The production dynamics is dominated by the $\omega(782)\pi^0\eta$ 
final state with  a  possible small contribution
of the $a_0(980)$ resonance in the $\pi^0\eta$ combination. We also
observe a presence of the $\rho(770)^{\pm}$ resonance signal in the
$\pi^{\pm}\pi^0$ invariant mass distribution.}

\begin{document}


\maketitle
\flushbottom


\baselineskip=17pt
\section{ \boldmath Introduction}
\hspace*{\parindent}
The $\epem$ annihilation into hadrons below 2\,GeV is rich in various
multiparticle final states. 
Detailed investigation of these states is important for the development of
phenomenological models describing strong interactions at low energies.
 The $\epem\to \pipi\ppz\eta$ cross section
contributes a not negligible value 
(reaching a maximum of up to 5\% of the total hadronic cross section)
to the calculations of the hadronic contribution to the muon anomalous magnetic 
moment~\cite{g-2}. A detailed study of the production dynamics can
further improve the accuracy of these calculations as well as our 
understanding of the spectroscopy of light mesons.

In the first study by the SND Collaboration~\cite{sndompi0eta} the cross section for the $\epem\to\omega(782)\piz\eta$ ($\omega\to\piz\gamma, \eta\to\gamma\gamma$) reaction was presented in the center-of-mass energy (\Ecm) below 2.0\,GeV using detection of seven photons.
Later on, the BaBar Collaboration  presented the inclusive $\epem\to\pi^+\pi^-\pi^0\pi^0\eta$ cross section~\cite{isr2pi6g} in a wide \Ecm~range
with many intermediate states demonstrated.
The $\epem\to\omega(782)\pi^0\eta$ reaction was confirmed by BaBar to dominate below \Ecm=2\,GeV.
The SND measurement was performed with relatively low statistical accuracy, and the cross sections obtained in these experiments were systematically different.
 The new measurement of the $\epem\to\pipi\ppz\eta$ cross section performed by SND~\cite{snd2pi6g} is in a reasonable agreement with the BaBar data.

In this paper we present the analysis of the
$\epem-\to\pi^+\pi^-\pi^0\pi^0\eta$ process based on 372 pb$^{-1}$ of
the integrated luminosity collected at the CMD-3 detector 
in the \Ecm=1600--2007\,MeV energy  range at  the VEPP-2000
collider~\cite{vepp}. The inclusive cross section and the cross
section for the dominant  $\epem\to\omega(782)\pi^0\eta$  reaction
are presented.

\section{The CMD-3 detector and data collection}
\hspace*{\parindent}
The general-purpose detector CMD-3 has been described in 
detail elsewhere~\cite{sndcmd3}. The detector tracking system consists of a 
cylindrical drift chamber (DC)~\cite{dc}
placed inside a thin (0.2~X$_0$) superconducting solenoid with a field of 1.3~T.
The tracking system allows for the detection of charged tracks with a minimum polar 
angle of about 0.5 radians relative to the beam axis (about 90\% of
4$\pi$) and provides signals for the trigger from charged particles.
The barrel liquid-xenon (LXe) calorimeter with a 5.4~X$_0$ thickness has
fine electrode structure, providing a 1--2 mm spatial resolution 
for photons~\cite{lxe}, and
shares the cryostat vacuum volume with the superconducting solenoid.     
The barrel CsI crystal calorimeter is placed outside  the LXe calorimeter 
and  increases the total thickness to   13.5~X$_0$.  The end cap BGO 
calorimeter with a thickness of 13.4~X$_0$ is placed inside the 
solenoid~\cite{cal}.
The combined calorimeter allows for the detection of   photons with a minimum polar 
angle down to 0.25 radians relative to the beam axis (about  98\% of
4$\pi$) and provides trigger for  neutral particles.
The luminosity is measured using events of Bhabha scattering 
at large angles with about 1\% accuracy~\cite{lum}. 

The beam energy and the energy spread have been monitored using 
the Back-Scattering-Laser-Light system~\cite{laser}.
The 0.1\,MeV accuracy in the average \Ecm~ value and the energy spread of
about 0.6--0.7\,MeV have practically no effect on the uncertainty in the cross
section measurement.

To understand the detector response to the processes under study and obtain the
detection efficiency, we have developed a Monte Carlo (MC) simulation of the
detector based on the GEANT4~\cite{geant4} package, with all generated events passing
the full reconstruction and selection procedure.
The MC simulation uses primary generators with matrix elements 
for the studied processes, including soft photon radiation by the initial 
electron or positron calculated according to ref.~\cite{kur_fad}.

For the background study, we have developed a special MC generator~\cite{eetohadrons} to
 generically simulate the $e^+e^-\to hadrons$ reaction, including the majority
 ($>30$) of the exclusive channels weighted by their known cross sections.

In this analysis we make use of the data collected at 42 energy points 
during three energy scans performed in 2020, 2021 and 2022.

%
%
\section{Selection of $\epem\to\pipi\ppz\eta$ events}
\label{select}
\subsection{Preliminary selections}
\hspace*{\parindent}
We identify the 
$\pipi\ppz\eta$ candidate events using the $\eta\to\gamma\gamma$ decay.
Candidates for the process under study are required to have two good tracks from
oppositely charged particles and six or more clusters in the calorimeters that
are not associated with the tracks and are therefore assumed to be produced by
photons.
Each track is required to have ionization losses in the DC consistent with the
pion hypothesis, a momentum greater than 40\,MeV/c, and a transverse
(longitudinal) distance from the beam axis (interaction region center) of less
than 0.25\,cm (12\,cm).
The photon candidate is required to have an energy deposition in the calorimeters 
exceeding 25 MeV.

The selection criteria are very loose and do not introduce 
significant systematic uncertainties, which are studied by variations
of the selections. We observe no candidate events below \Ecm~= 1600 MeV. 

\begin{figure}[tbh]
\begin{center}
\includegraphics[width=0.49\textwidth]{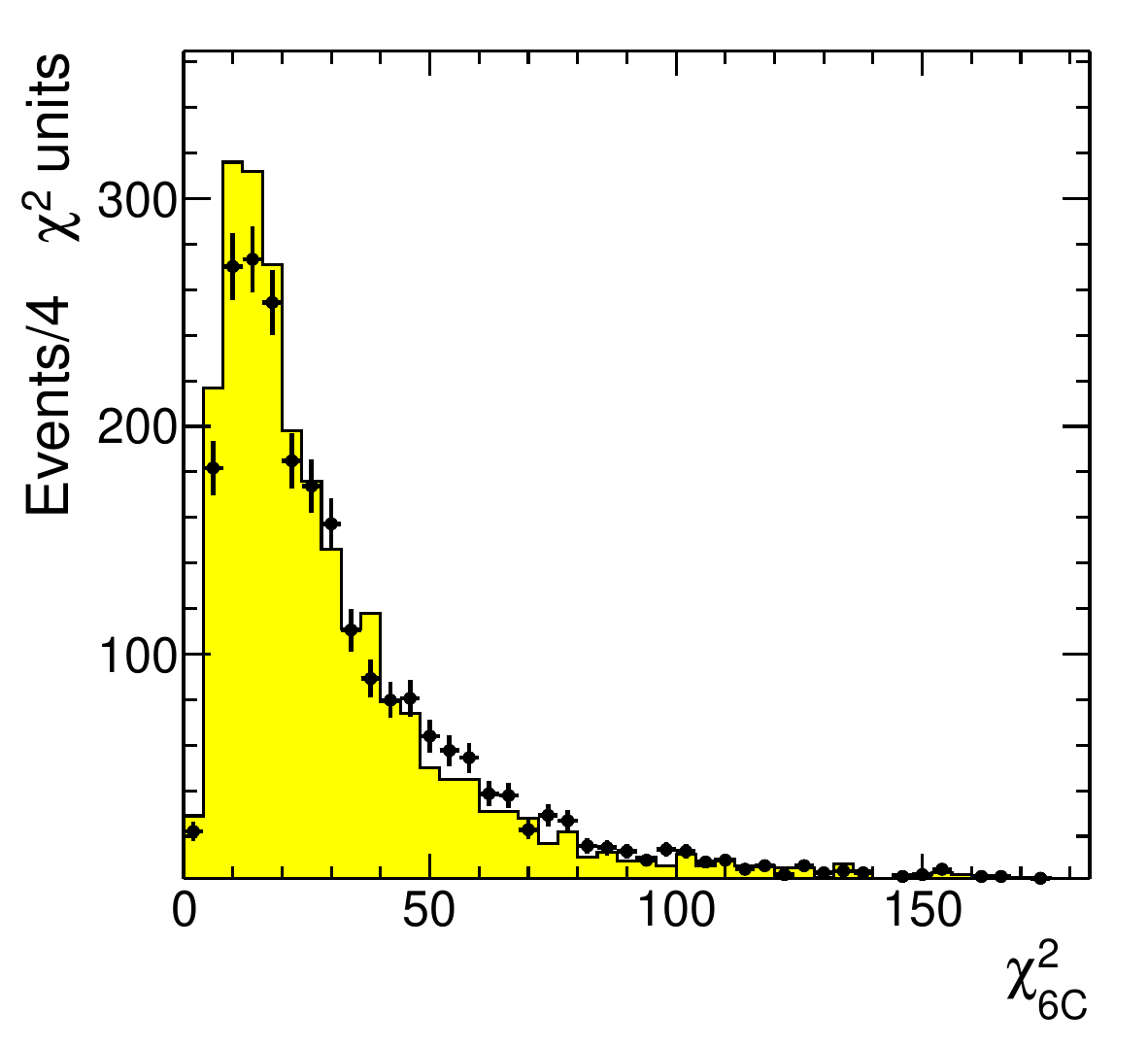}
\put(-50,140){\makebox(0,0)[lb]{\bf(a)}}
\includegraphics[width=0.46\textwidth]{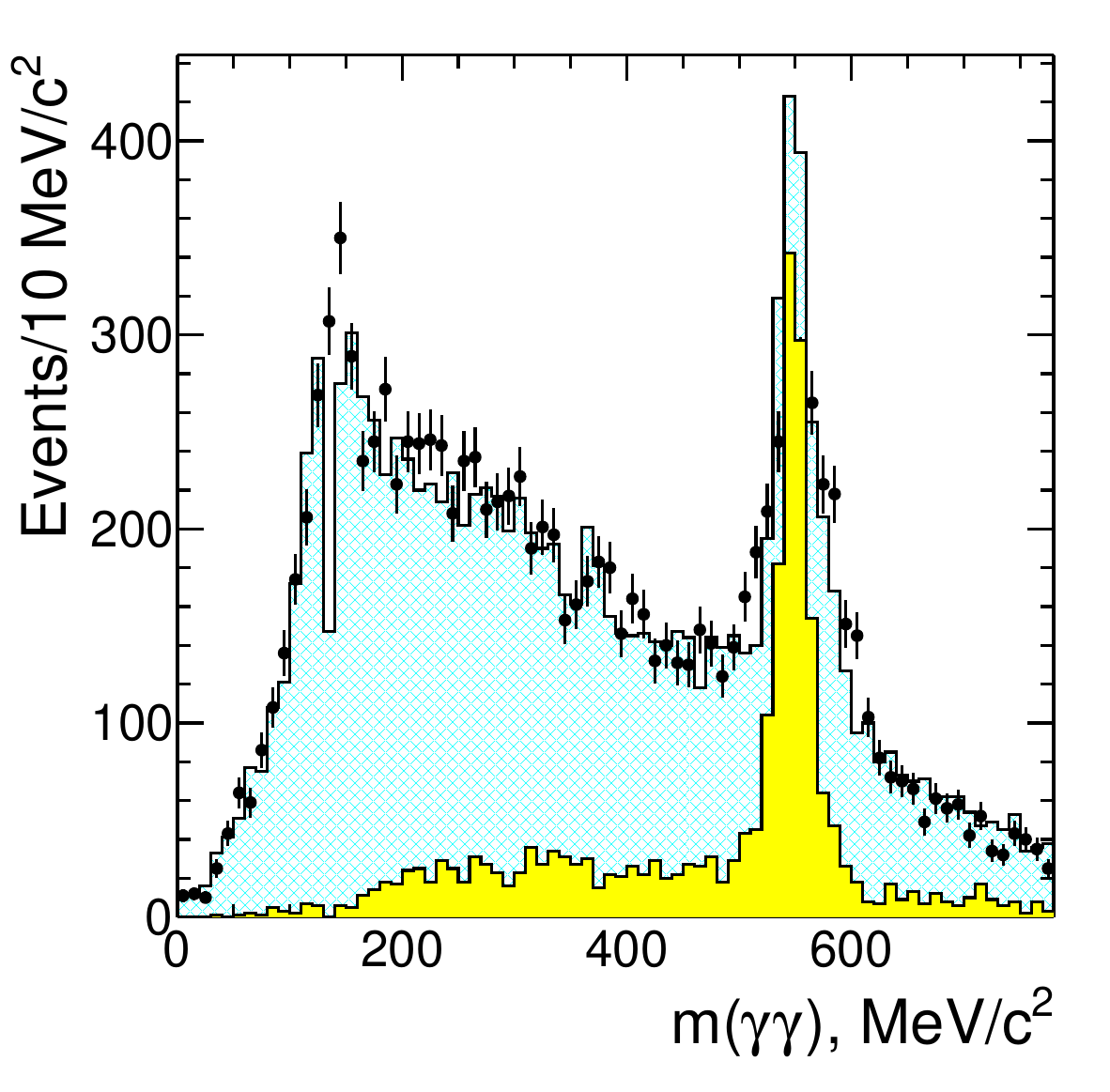}
\put(-50,140){\makebox(0,0)[lb]{\bf(b)}}
\vspace{-0.3cm}
\caption
{
(a) The 6C-fit $\chi^2$ distribution for events with two tracks, two \piz,
and two photons under the $\epem\to\pipi\ppz\gamma\gamma$ hypothesis
for data (points) and the corresponding simulation (histogram). (b) The
experimental invariant mass distributions of the unconstrained photon pair 
before (points) and after (cross-hatched histogram) the kinematic fit, with the corresponding simulated (shaded) histogram.
}
\label{chi2}
\end{center}
\end{figure}

\subsection{Kinematic fit}
\hspace*{\parindent}
The reconstructed momenta and angles of the detected charged tracks as well as 
the energy and the angles of the six photons are subject to the kinematic fit for 
the $\epem\to\pipi\ppz\gamma\gamma$ hypothesis,
with the total energy and momentum constrained by \Ecm and zero,
respectively.
We use the kinematic fit package described in ref.~\cite{kinfit}.
For each combination of the six photons in an event, we consider fifteen independent groupings of these photons into three pairs, requiring the invariant masses of at least
two pairs in the grouping to be close to the \piz~mass within a
$\pm$65\,\mevcc~(about $\pm$3.5~standard deviations) window. The
invariant masses of these two photon pairs are constrained to the \piz~mass
in the kinematic fit (6C~fit), leaving the third pair unconstrained. If all three photon pairs in the grouping satisfy
the \piz~window condition, we search for the alternative with the lowest
$\chi^2$ among all three options.
The co-variance matrices for the  charged tracks and the photons are
used in the fit and provide a $\chi^2$ value for each event for each
combination.

A large fraction of the event candidates contains more than six photons. All
possible combinations are tested, and the three photon pairs with the smallest
$\chi^2$ value are retained for further analysis. As a result of the fit, we obtain   
improved values of the momenta, energies, and angles for all particles.

Figure~\ref{chi2}(a) shows the obtained $\chi^2$ distributions for the
experimental (points) and simulated $\epem\to\pipi\ppz\eta$ (histogram)
events, where the invariant mass of the unconstrained photon pair is in the 
$\pm$70~\mevcc~window around the $\eta$ mass. 
The requirement $\chi^2_{\pi^+\pi^-\pi^0\pi^0\gamma\gamma}<130$ is applied as a
selection criterion for the signal events.

Each event is also subject to the 5C kinematic fit under the
$\epem\to\pipi\piz\gamma\gamma$ hypothesis: all photon pairs are tested to get 
the best $\chi^2$ value, and a requirement $\chi^2_{\pipi\piz\gamma\gamma}>20$ 
suppresses the background from the processes $\epem\to\pipi\ppz$ and 
$\epem\to\pipi\piz\eta$ by a factor of 10--20 to a negligible level with a 
3\% loss of the signal events.

Figure~\ref{chi2}(b) demonstrates the invariant mass distributions for the 
unconstrained photon pair before (points) and after (cross-hatched  histogram) the 
6C kinematic fit for events in the \Ecm= 1970-2007~MeV energy range 
with the applied $\chi^2<130$ selection. The corresponding simulated events are shown by a shaded histogram.
A signal from the $\eta\to\gamma\gamma$ decay 
is clearly seen, and an improvement in the resolution is obtained.

A dip near the \piz~mass in the unconstrained photon pair invariant mass
distribution arises because a photon pair with an invariant mass 
close to the \piz~mass is highly likely to be captured by the mass constraint.
Selection of the best combination from the three tested around the  \piz~mass moves two
best photon pairs under the constrains, and leaves a dip for the third pair invariant mass. 

The  background from other multi-pion processes is relatively large.
To study the remaining background, we analyze 
events from the generic $\epem\to hadrons$ MC
generator~\cite{eetohadrons} with the signal process excluded, and
find a very low additional background with no events producing a peaking background under the $\eta$ signal.
\begin{figure}[tbh]
\begin{center}
\includegraphics[width=0.49\textwidth]{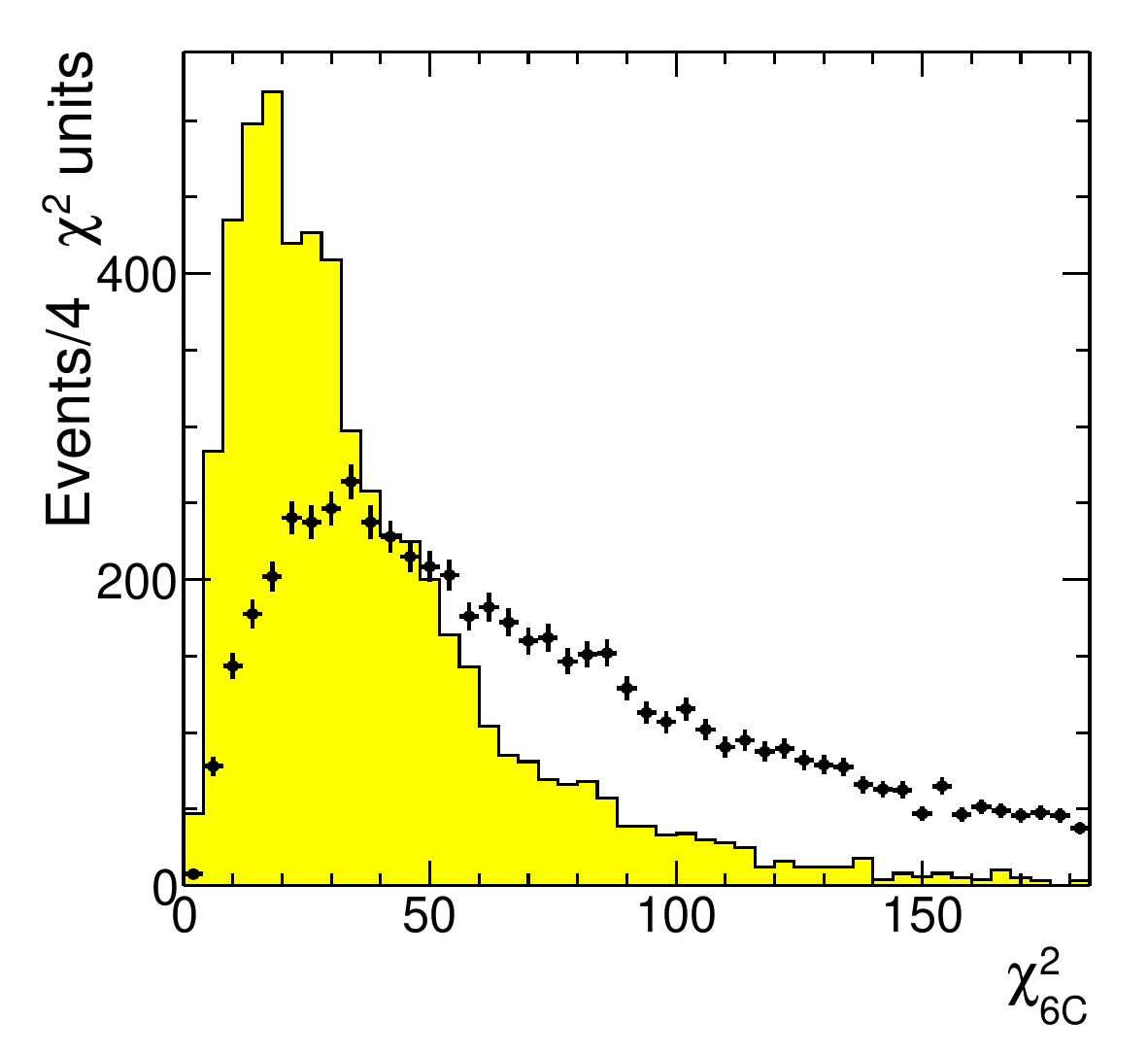}
\put(-50,140){\makebox(0,0)[lb]{\bf(a)}}
\includegraphics[width=0.49\textwidth]{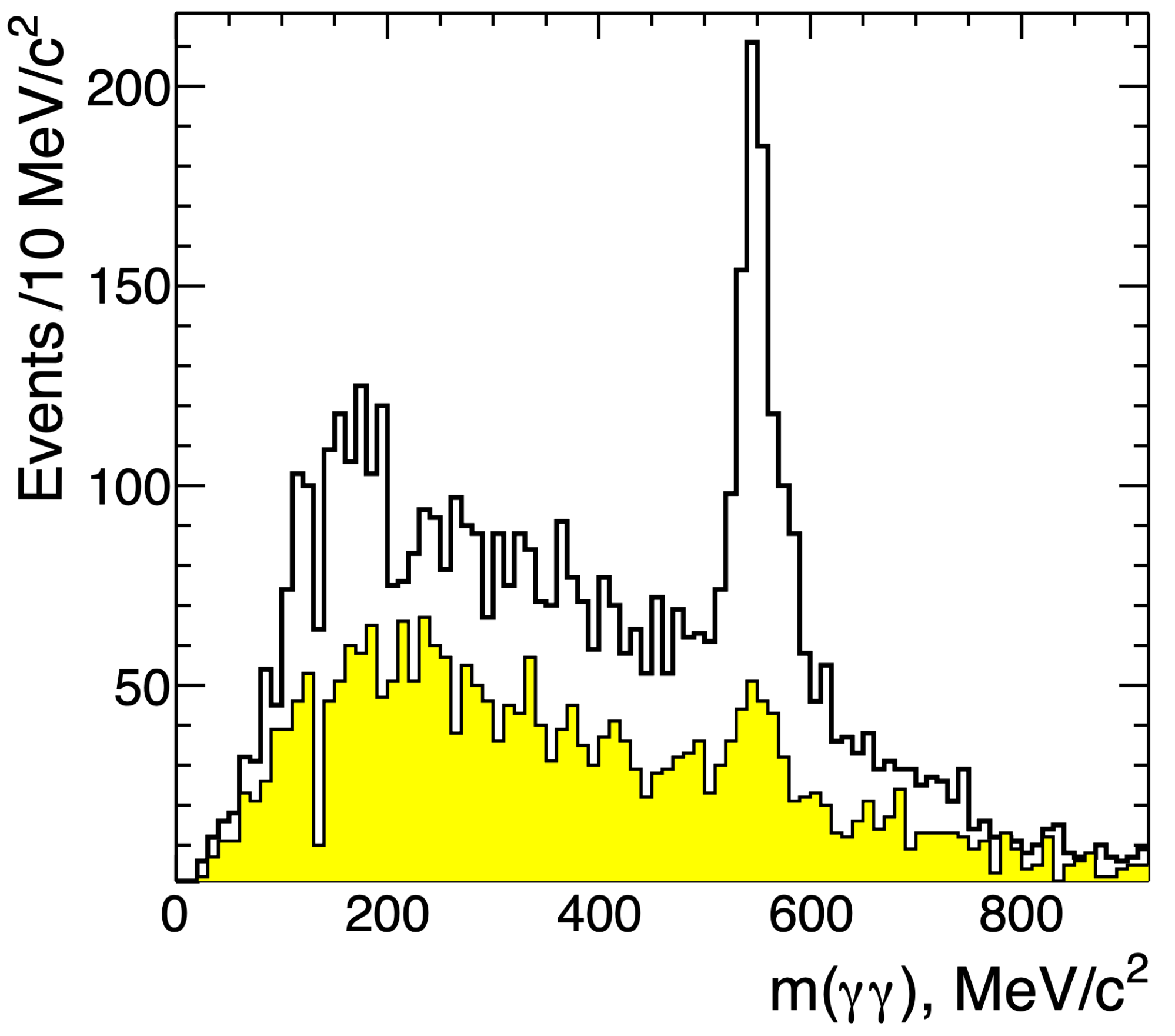}
\put(-50,140){\makebox(0,0)[lb]{\bf(b)}}
\vspace{-0.3cm}
\caption
{
(a) The 6C-fit $\chi^2$  distributions for the events with two tracks, two \piz, and two photons
for the $\epem\to\pipi\ppz\gamma\gamma$
hypothesis for the best $\chi^2$ value (shaded histogram) and for the
second-best value (points). (b) The experimental two-photon invariant mass distributions
for the best $\chi^2 <130$ value selection (open histogram) and for
the same selection using the second-best $\chi^2$ value (shaded histogram).
}
\label{chi2_2nd}
\end{center}
\end{figure}

\subsection{Combinatorial problem}
\hspace*{\parindent}

A large fraction of the  events is observed to have
an acceptable  $\chi^2 < 130$ value for another combination of 
the same six photons.
This second-best $\chi^2$ value is slightly worse, but
the corresponding combination can contain the  
$\eta\to\gamma\gamma$ signal instead of the best one.  
Because of the limited photon energy resolution,  the kinematic fit sometimes
finds  a wrong  combination of three pairs with the best  $\chi^2$ value.
Figure~\ref{chi2_2nd}(a) shows the $\chi^2$ distributions 
for the photon combination with the best  $\chi^2$ value (histogram) and the 
distribution for the alternative combination of three pairs having an acceptable
$\chi^2$ value (points).  

Figure ~\ref{chi2_2nd}(b) shows the $\gamma\gamma$ invariant mass distribution for
the events selected by the best $\chi^2<130 $ value (open histogram), and for the same 
events which have the second-best acceptable $\chi^2<130$ value (shaded histogram).
An additional $\eta\to\gamma\gamma$ signal is observed in both  data and simulation 
for the second-best $\chi^2$ value, while no signal is present for the best one. 
If the $\eta$ signal exists in the combination with the second-best $\chi^2<130$
value, the difference between this value and the best one does not exceed five to six
units; we use this property for further combinatorial background suppression.

This effect can differ between data and simulation.
To minimize a possible difference in the number of events, we use the sum of
 two distributions: the first is obtained by requiring the best $\chi^2<130$,
while the second includes events where the second-best $\chi^2$ value differs by
 no more than six units from the best one.
An  example of these two distributions is shown in figure~\ref{chi2_2nd}(b): the
sum of the appropriate distributions is used in all further
analysis. 

We also check that if an event has the second-best $\chi^2<130$ for a different
set of six photons, no signal contribution is observed.
The  number of events with the third-best $\chi^2<130$ value is relatively small , and these events do not exhibit a sizeable  signal.

\begin{figure}[tbh]
\begin{center}
\includegraphics[width=1.03\textwidth]{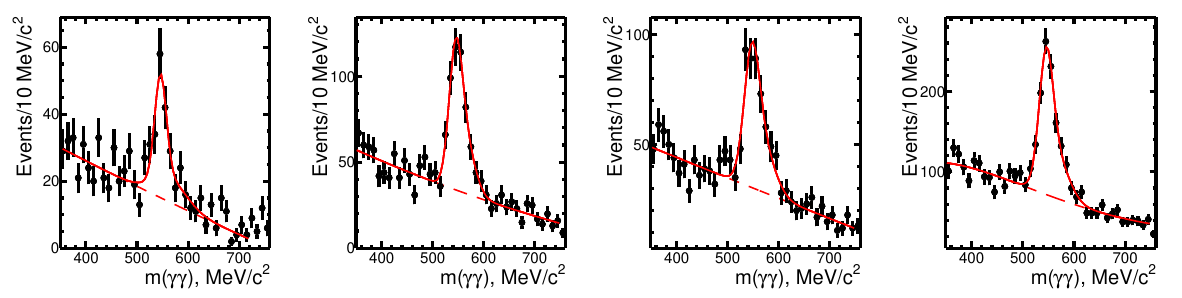}
\vspace{-0.3cm}
\caption
{
Examples of the two-photon invariant mass distributions and fit functions to determine 
the number of the $\pipi\ppz\eta$ events at \Ecm = 1970, 1980, 1990,
and 2007 MeV.
}
\label{etafit}
\end{center}
\end{figure}

\subsection{Signal event extraction}
\hspace*{\parindent}
The $\eta$ peak in the invariant mass distribution of the third photon pair 
is used to obtain the inclusive number of $\pipi\ppz\eta$ events. We fit the sum of 
the distributions in figure~\ref{chi2_2nd}(b) at each energy point with a sum of functions
to separate the signal and the background events. A double-gaussian function is used
for the $\eta$ signal, while a second-order polynomial function describes
the background. The example of the fits is shown in
figure~\ref{etafit} for the points from the 2021 scan: \Ecm = 1970,
 1980, 1990, and 2007 MeV.
The total number of the $\pipi\ppz\eta$ events evaluated by this
procedure is 6300$\pm$145 and listed in table~\ref{xs_all}. We do not
observe any signal  events for \Ecm below 1600~MeV, and present our
data starting  from \Ecm=1600 MeV.  

A variations of the signal and background shapes do not change number
of the events by more than  3\%.

The observed $\pipi\ppz\eta$ events contain several intermediate 
states. Our data sample is too small for a standard amplitude analysis.
Instead, we first extract the contribution of the dominant $\omega(782)$
intermediate resonance, and then investigate other contributions, 
assuming low interference with the narrow state above.

\begin{figure}[tbh]
\begin{center}
\vspace{-0.2cm}
\includegraphics[width=0.49\textwidth]{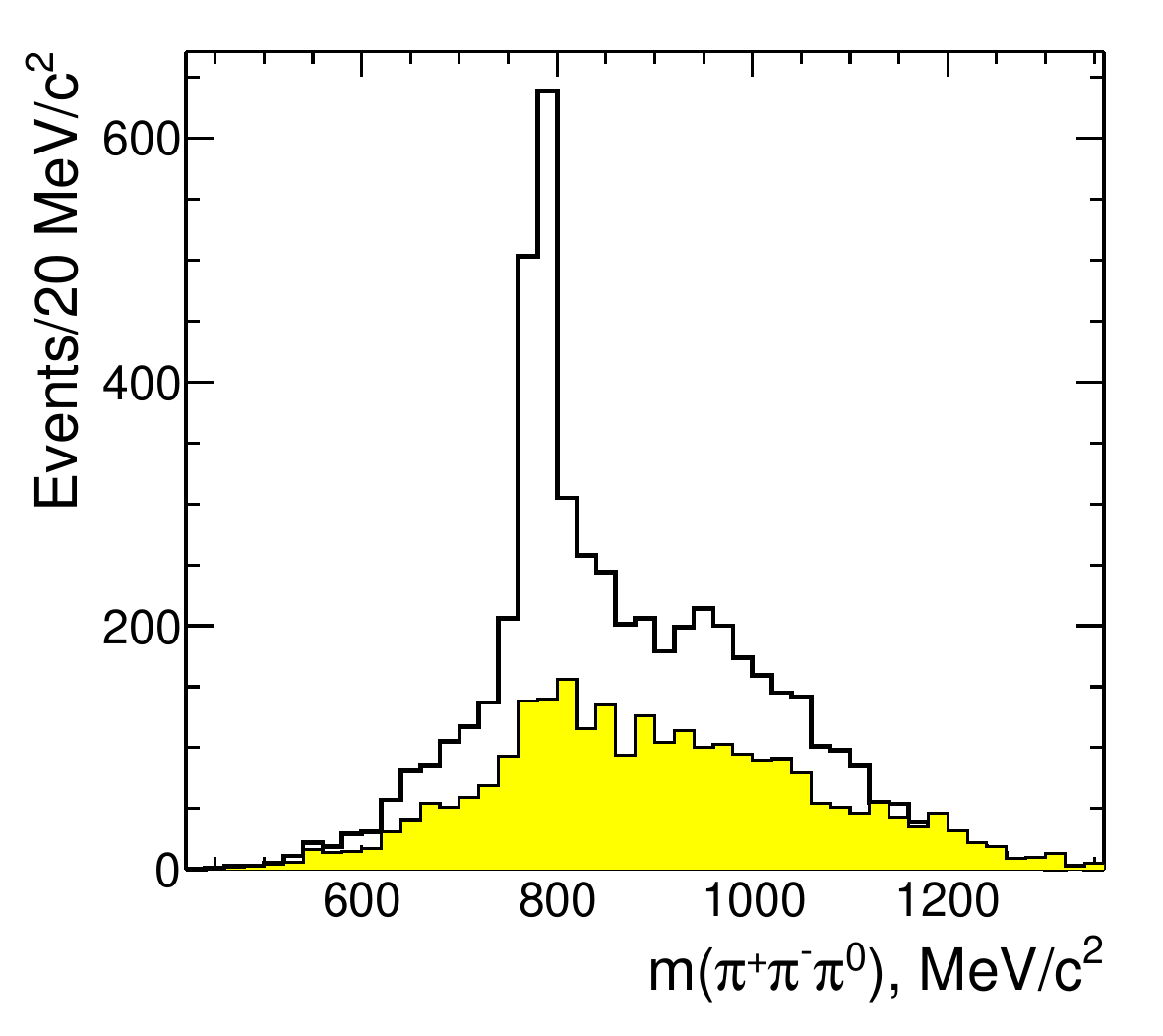}
\put(-50,140){\makebox(0,0)[lb]{\bf(a)}}
\includegraphics[width=0.49\textwidth]{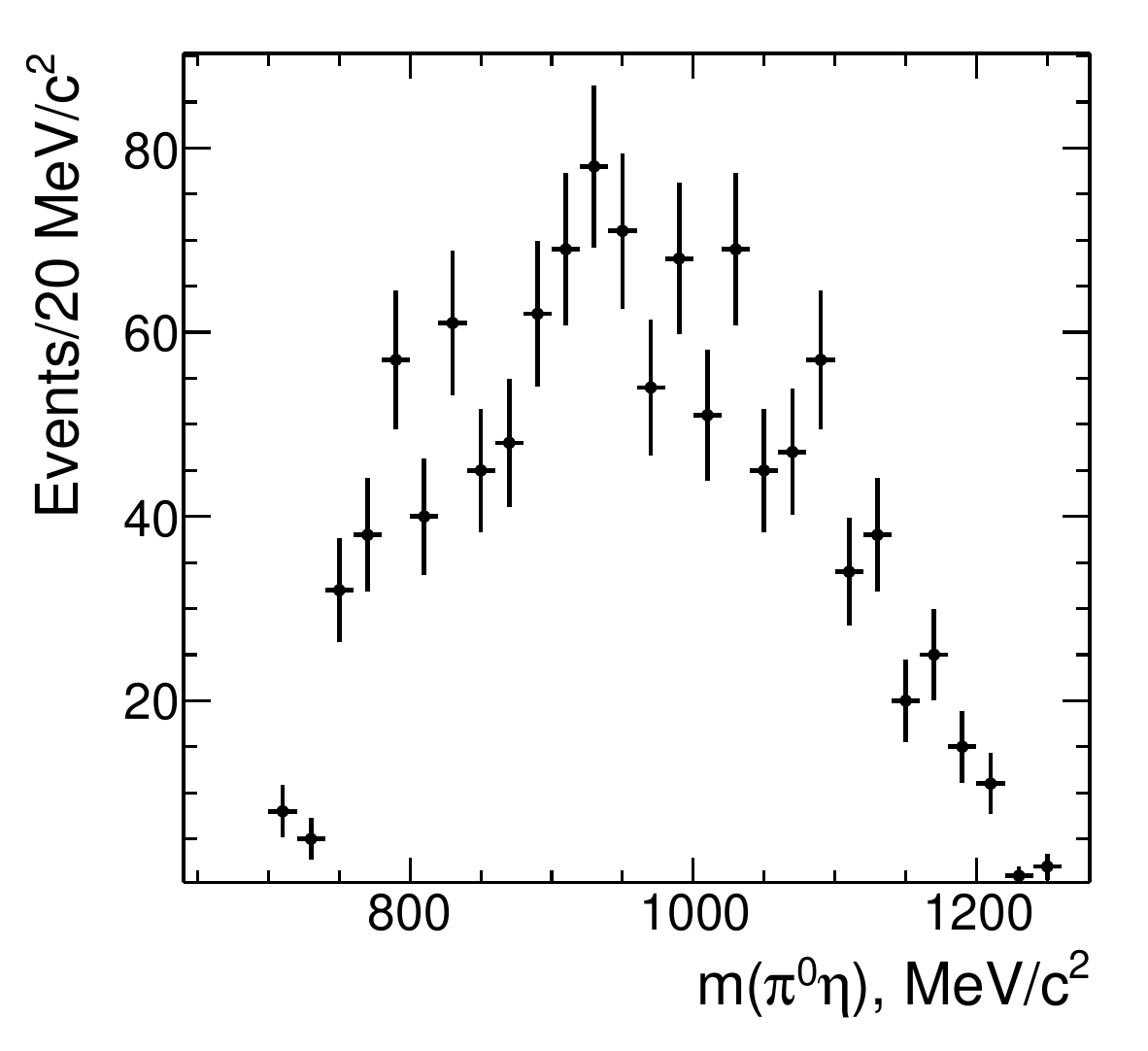}
\put(-50,140){\makebox(0,0)[lb]{\bf(b)}}
\vspace{-0.3cm}
\caption{
(a) The three-pion invariant mass (two combinations) from the events in
the signal region of figure~\ref{chi2}(b) (open histogram) and side band region
(shaded histogram).
(b) The background subtracted $\piz\eta$ invariant mass for the events
from the $\omega$ peak of (a).
}
\label{3pimass}
\end{center}
\end{figure}

\section{The $e^+e^-\to \omega(782)\piz\eta$ intermediate state}
\label{xsomegaeta}
\hspace*{\parindent}
To study intermediate states we select signal candidates by requiring
 $|m_{\gamma\gamma}-m_{\eta}| < 70$~\mevcc, see figure~\ref{chi2}(b), and subtract 
the side band background using events with  
$70 < |m_{\gamma\gamma}-m_{\eta}| < 140$~\mevcc ~for all experimental distribution.
Figure~\ref{3pimass}(a) shows the $\pipi\piz$ 
invariant mass distributions (two combinations) for the selected 
$\pipi\ppz\eta$ candidates from the signal region (open histogram) and
from the $\eta$ side band regions, shown by a shaded histogram.
A remaining peaking background  is seen, dominated by the
$\epem\to\omega\piz$ reaction with large cross section.
The signal from the $\omega(782)$ dominates at all energy points, but a possible
signal from the $\phi(1020)\to\pipi\piz$ resonance is not observed. 

To determine number of the $\omega$ events, we subtract the events
from the side band region, and  fit the distributions  
at each energy with a sum of the signal and combinatorial background
functions. For the signal peak we use the Breit-Wigner function,
convolved with an additional Gaussian function to take into account
the detector resolution (about 10-12~\mevcc). 
A smooth polynomial function is used to 
describe the combinatorial background. 
  
\begin{figure}[tbh]
\begin{center}
\includegraphics[width=1.03\textwidth]{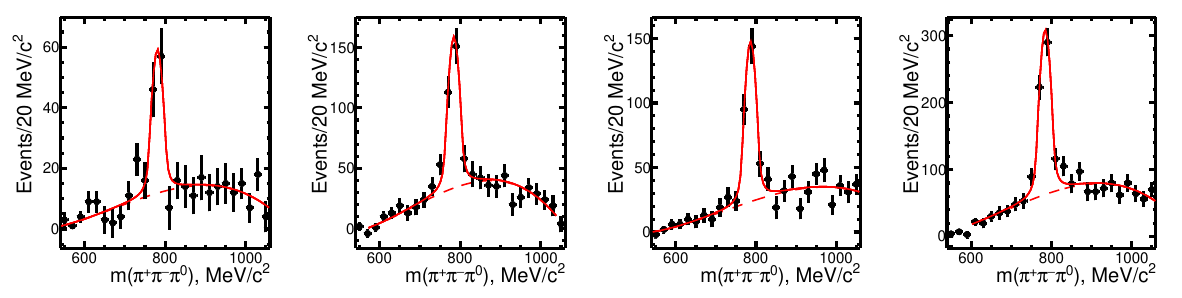}
\vspace{-0.3cm}
\caption
{
Example of the background subtracted three-pion invariant mass
distributions and fit functions to determine  
the number of the $\omega\piz\eta$ events at \Ecm = 1970, 1980, 1990,
and 2007  MeV.
}
\label{omegafit}
\end{center}
\end{figure}

Histograms in figure~\ref{omegafit} show examples of the fit
to the $\omega(782)$ signal for the \Ecm = 1970, 1980,1990, and 2007 MeV energy points.
 
In total, for all energy points we obtain 
$6024\pm119$ events for the $\omega(782)\piz\eta$ intermediate state. 
This number is consistent with that obtained for the inclusive
$\epem\to\pipi\ppz\eta$ reaction, indicating a dominance of the $\epem\to\omega\piz\eta$ reaction.

By varying  the 
polynomial order of the background function or by removing the side band 
background subtraction, we estimate the systematic uncertainty on the number 
of signal events of about 7\%. 

\begin{figure}[tbh]
\begin{center}
  \vspace{-0.2cm}
  \includegraphics[width=0.33\textwidth]{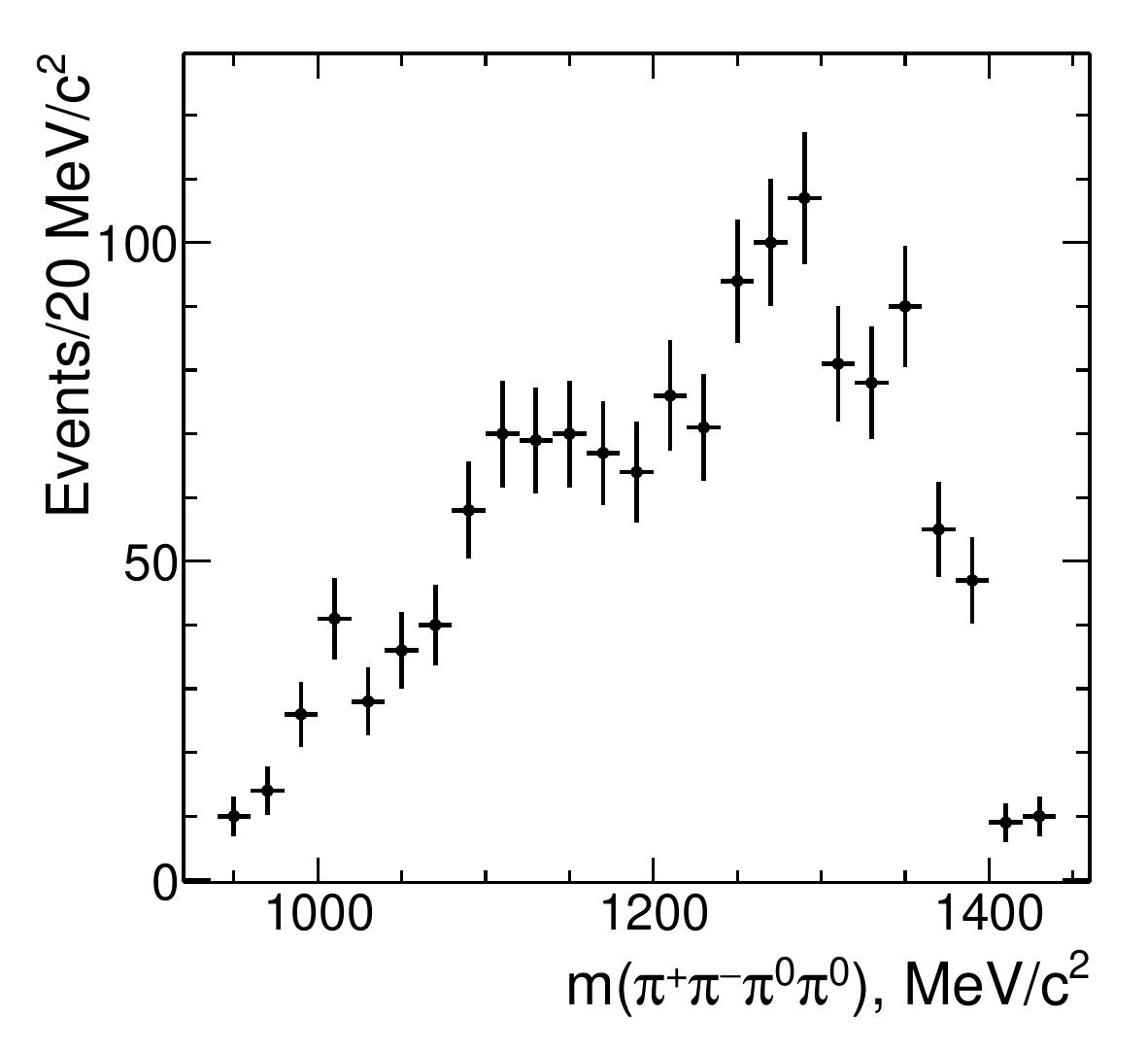}
\put(-100,100){\makebox(0,0)[lb]{\bf(a)}}
\includegraphics[width=0.33\textwidth]{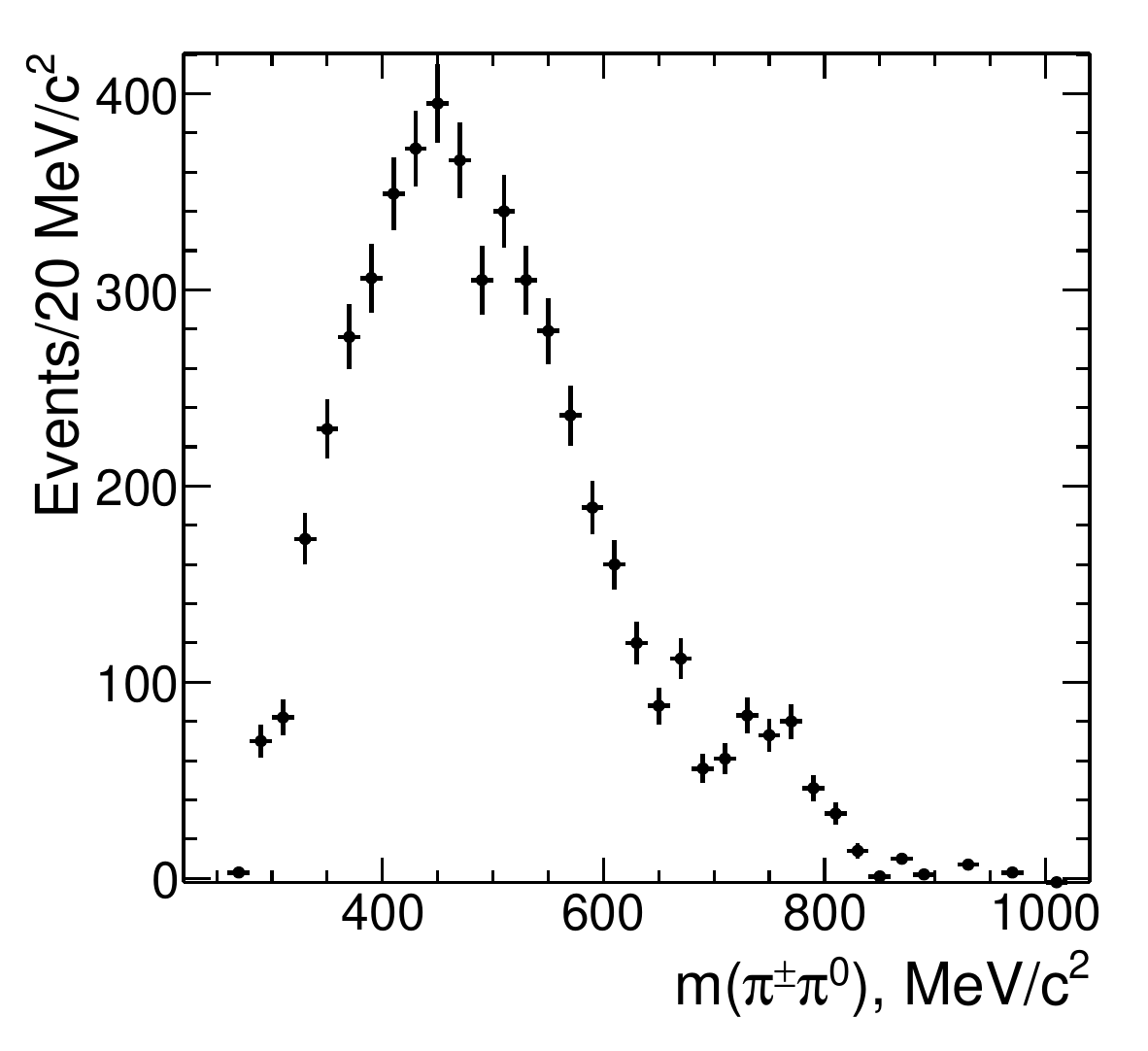}
\put(-50,100){\makebox(0,0)[lb]{\bf(b)}}
\includegraphics[width=0.33\textwidth]{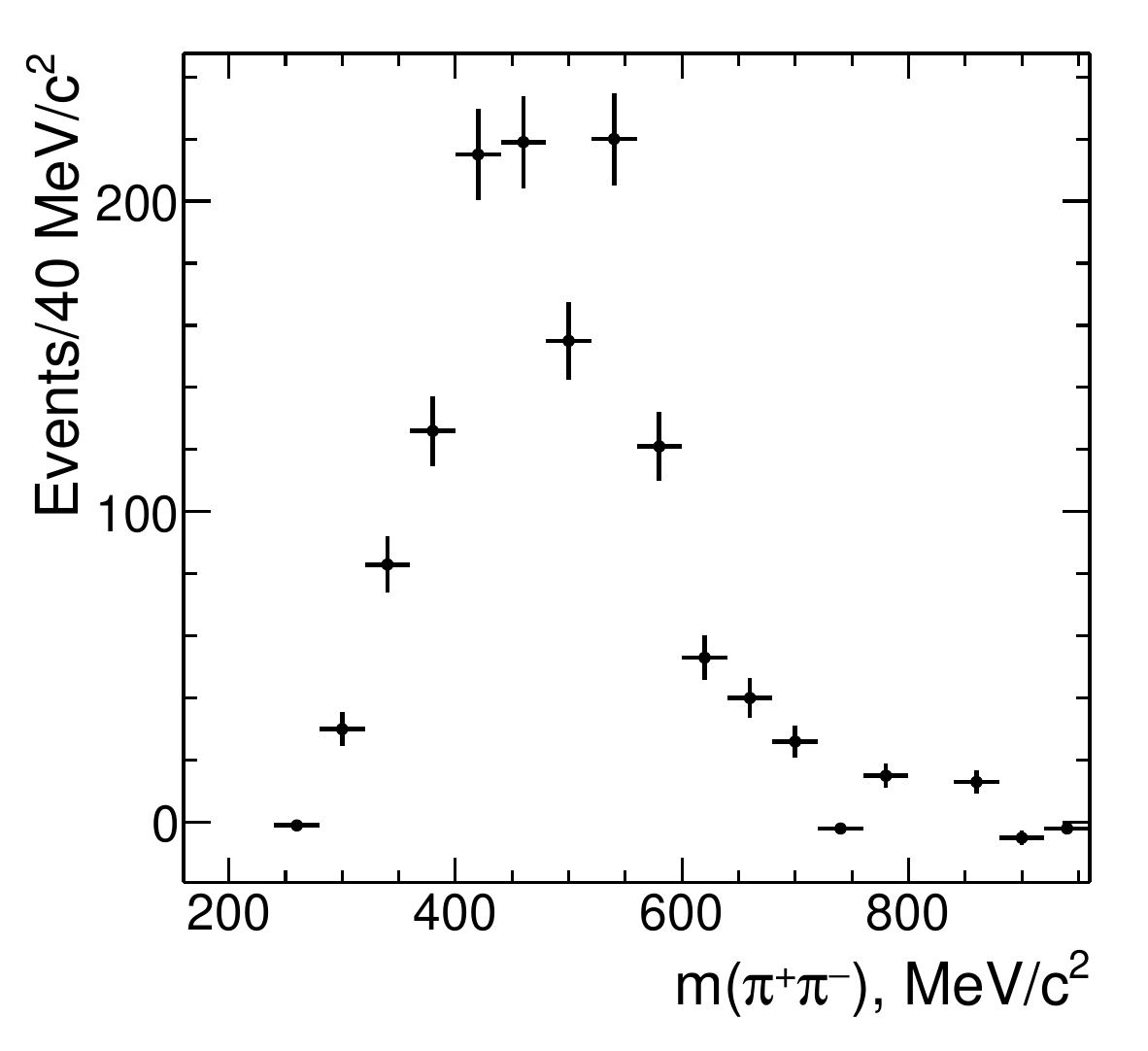}
\put(-50,100){\makebox(0,0)[lb]{\bf(c)}}
\vspace{-0.3cm}
\caption{
The background-subtracted 
$\pipi\ppz$ (a),  $\pi^{\pm}\piz$ (b), and $\pipi$ (c)  invariant 
mass distributions for the events in the 1970--2007 MeV energy range. 
}
\label{mpieta}
\end{center}
\end{figure}
\section{Search for other intermediate states}
\label{xsa0rho}
\hspace*{\parindent}
As seen in figure~\ref{omegafit}, the combinatorial background for the
$\omega(782)\piz\eta$ final state is  a little
larger than expected, and other intermediate resonances
could contribute: the most probable are
$a_0(980)\omega(782), (a_0(980)\to\piz\eta)$ and
$\rho(1450,1700)\eta, (\rho\to\pipi\ppz)$.  
The $a_0(980)$ is relatively narrow (about 50 MeV~\cite{pdg}) and
should be seen in the $\eta\piz$ invariant mass. 
Figure~\ref{3pimass}(b) shows the background-subtracted 
$\piz\eta$ invariant mass distribution 
for the events in the \Ecm= 1970--2007 MeV energy range with
additional requirement for the three-pion mass to be in the
$\pm$40~\mevcc~ window around the $\omega$ mass. 
No signal associated with the $a_0(980)$ resonance is observed in our energy range, 
while the $a_0(980)\omega$ intermediate state is reported by
BaBar~\cite{isr2pi6g} at higher energies. 

Figure~\ref{mpieta}(a) shows the background-subtracted 
$\pipi\ppz$ invariant mass distribution, demonstrating a structure 
from a possible intermediate $\epem\to\rho(1450,1700)\eta$ reaction.

 We also observe a clear signal from the $\rho^{\pm}(770)$ in the 
$\pi^-\piz,~\pi^+\piz$ corresponding mass combinations, 
shown in figure~\ref{mpieta}(b) for the \Ecm= 1970--2007 MeV range.
But no signal from the $\rho^{0}(770)$ in the $\pipi$ invariant mass
is observed in figure~\ref{mpieta}(c).

The contribution from the  intermediate states listed above  is relatively
small with the large combinatorial background from the dominant
$\epem\to\omega\piz\eta$ reaction, therefore, we do not extract the
number of events for these states.  
\begin{figure}[tbh]
\begin{center}
\vspace{-0.2cm}
\includegraphics[width=1.03\textwidth]{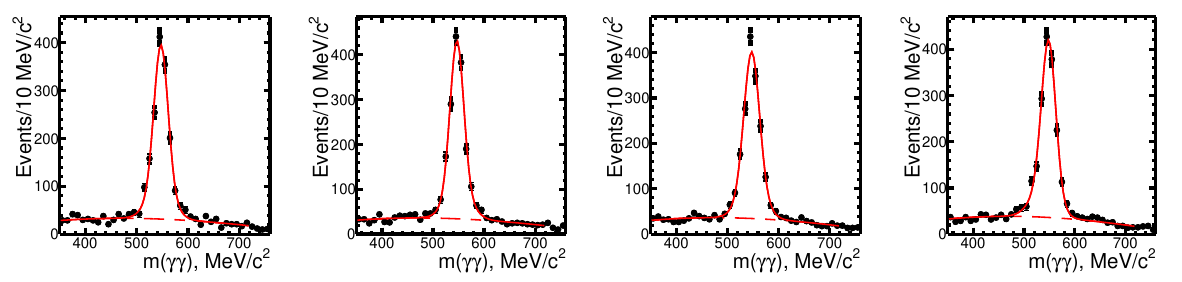}
\vspace{-0.3cm}
\caption
{
The  $\gamma\gamma$ invariant mass distribution for the events in the
\Ecm = 1970--2007 MeV energy range for simulation with the fit functions.
}
\label{m2gmassMC}
\end{center}
\end{figure}
\begin{figure}[tbh]
\begin{center}
\vspace{-0.2cm}
\includegraphics[width=1.03\textwidth]{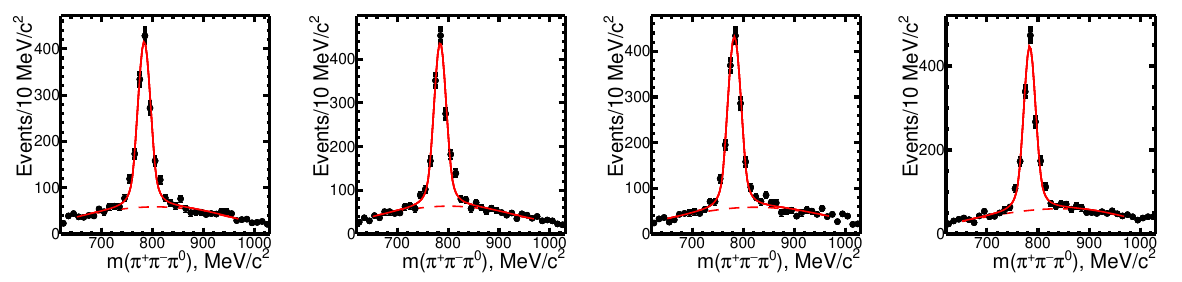}
\vspace{-0.3cm}
\caption
{
The  $\pipi\piz$ invariant mass distribution for the events in the
\Ecm = 1970--2007 MeV energy range for simulation with the fit functions.  
}
\label{3pimassMC}
\end{center}
\end{figure}
\begin{figure}[tbh]
\begin{center}
\vspace{-0.2cm}
\includegraphics[width=0.51\textwidth]{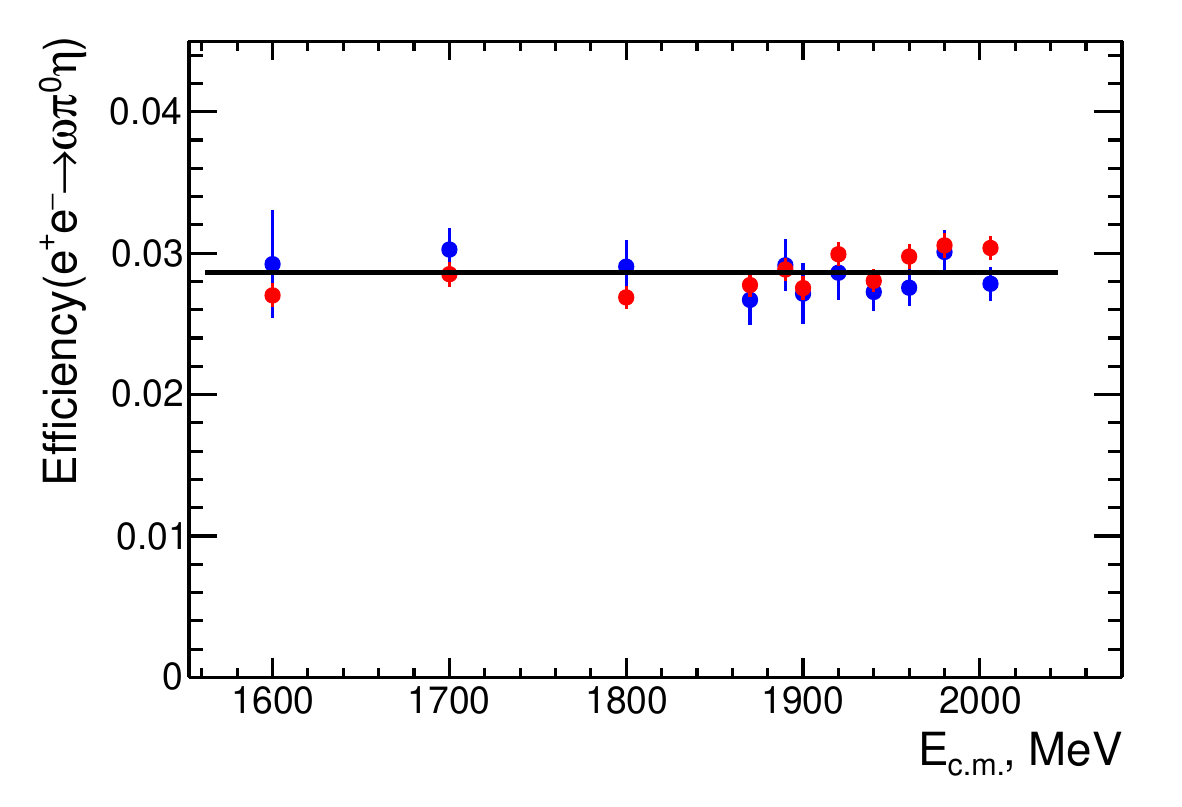}
\put(-50,120){\makebox(0,0)[lb]{\bf(a)}}
\includegraphics[width=0.51\textwidth]{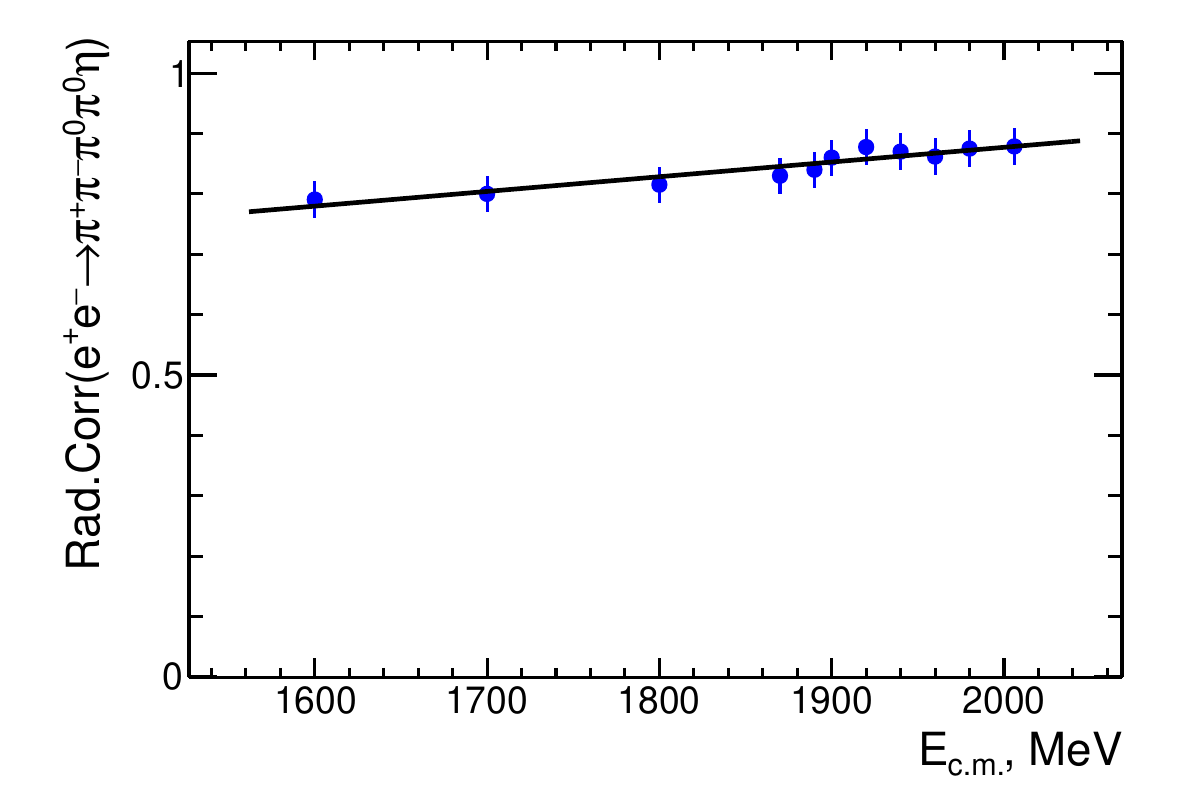}
\put(-50,100){\makebox(0,0)[lb]{\bf(b)}}
\vspace{-0.3cm}
\caption
{
(a) MC-calculated efficiency for the 
$\epem\to\omega(782)\piz\eta$ reaction determined from the $\eta$ peak
(red) and from the $\omega$ signal (blue).  
(b) Radiative correction ($1+\delta_R$) for the 
$e^+e^-\to \pipi\ppz\eta$ cross section.
}
\label{eff2pi6g}
\end{center}
\end{figure}
\section{Detection efficiency}
\label{eff}

To study the detector responce and efficiency we use about ten times
larger sample of the MC simulated events. The simulation uses the
primary generator for the $\epem\to\pipi\ppz\eta$ process,  producing events in a phase
space mode,  and in the mode with the $\omega(782)\piz\eta$ or  $\omega a_{0}(980)$
intermediate states. The generator includes a radiative photon
emission from the initial particles according to ref.~\cite{kur_fad}. 

The simulated events pass all above selections,  and
figure~\ref{m2gmassMC} shows an example of the $\eta$ signal extraction for simulation
similar to that for data in figure~\ref{etafit}. Figure~\ref{3pimassMC}
shows example for the $\omega$ signal extraction.

The $\epem\to\pipi\ppz\eta$~ simulation in the phase space
mode is not relevant to our case, and we do not use it for the
efficiency calculation. The other two modes of the primary generator produce
very close results.
 Figure~\ref{eff2pi6g}(a) shows the MC-simulated 
$\epem\to\pipi\ppz\eta$ detection efficiency, $\epsilon$, for the  $\omega(782)\piz\eta$
production mode determined as a ratio of events that pass reconstruction 
and selection criteria, to the total number of the simulated events. We
calculate the efficiency using the $\eta$ peak signal, or using the
$\omega$ signal indicated by color in
figure~\ref{eff2pi6g}(a). A relatively good agreement of these two
efficiencies indicates that the photons are correctly combined into $\piz$
and $\eta$ in the kinematic fit procedure. Note, that by adding the
event distribution from the second-to-best $\chi^{2}$ combination the
efficiency increases by about 20\%.

A fit with the constant yields the value $\epsilon = $0.286 which is used in the cross section calculation.
We do not observe more than 5\% difference for the different production modes or
for the different efficiency determination procedure, and use this
number as an estimate of the efficiency calculated systematic uncertainty.

\begin{figure}[tbh]
\begin{center}
\vspace{-0.3cm}
\includegraphics[width=1.0\textwidth]{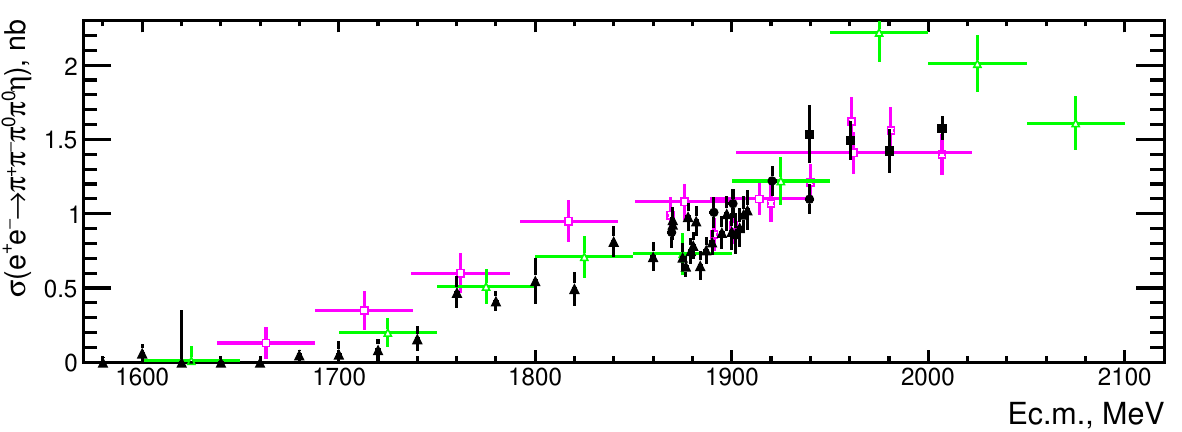}
\put(-395,64){\includegraphics[width=0.29\textwidth]{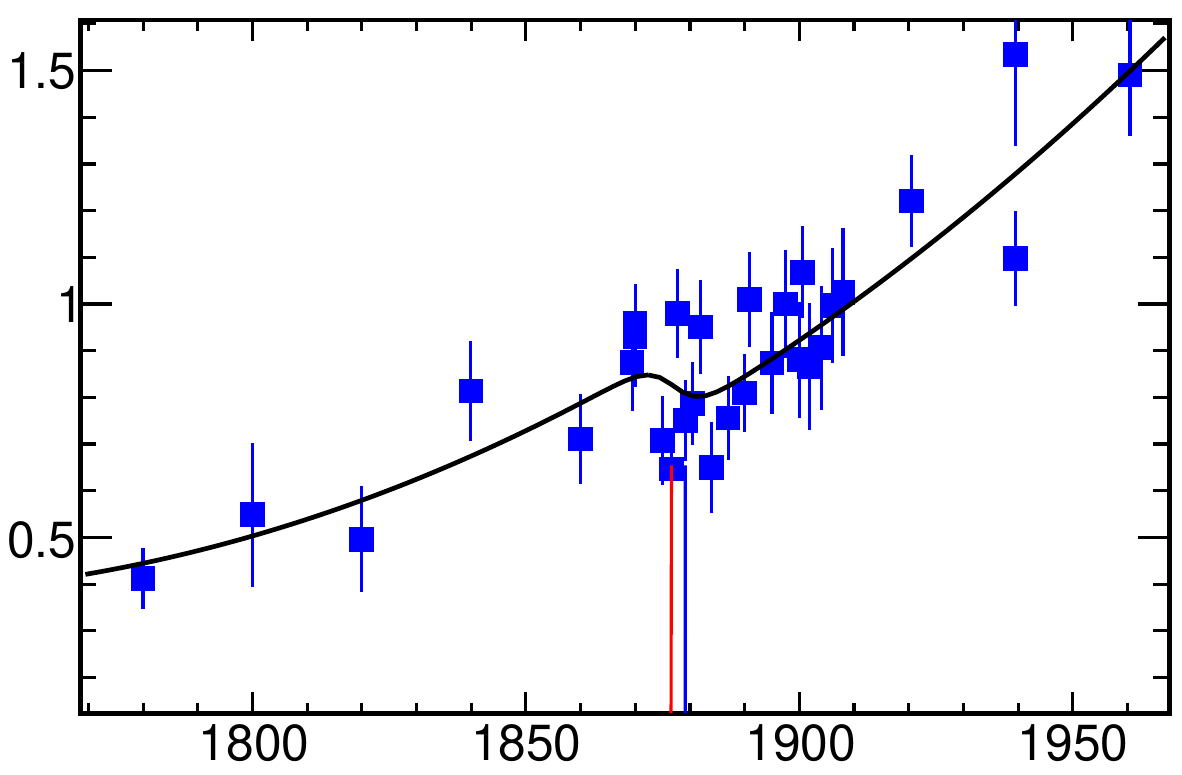}}
\put(-60,50){\makebox(0,0)[lb]{\bf(a)}}\\
\includegraphics[width=1.0\textwidth]{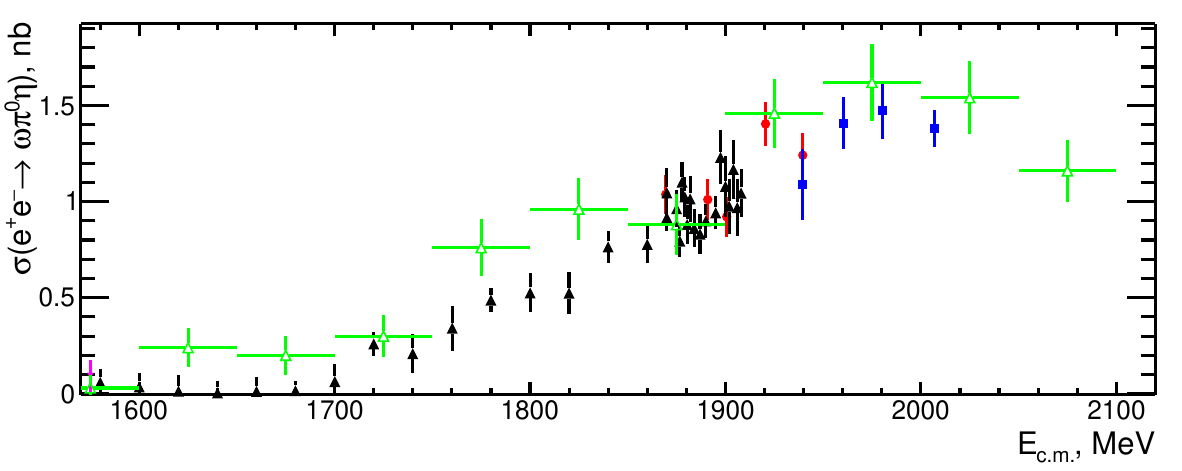}
\put(-60,50){\makebox(0,0)[lb]{\bf(b)}}
\vspace{-0.5cm}
\caption
{
The $e^+e^-\to \pipi\ppz\eta$ (a) and  $e^+e^-\to \omega\piz\eta$ (b) cross sections obtained with the CMD-3 detector. 
In the insert a region of the $N\bar N$ production 
threshold is shown with the fit function, described in the tex.     
Filled triangle, circle and square markers are for the
2022, 2020, and 2021 data, respectively.
Open triangle and square markers are for the BaBar~\cite{isr2pi6g} and SND~\cite{snd2pi6g} data, respectively.
}
\label{xs_2pi2pi0eta}
\end{center}
\end{figure}
\section{The cross section calculation}
\label{xs2pipi0eta}

The cross section for the $e^+e^-\to \pipi\ppz\eta$ process is
calculated for each energy point as
\begin{equation}
\label{xsformular}
\sigma(\pipi\ppz\eta) = \frac{N}{L\cdot\epsilon\cdot 
(1+\delta_R)\cdot\epsilon_{\rm corr}}~,
\end{equation}
where $N$ is the number of selected events, $L$ is the integrated luminosity, 
$\epsilon$ is the detection efficiency shown in figure~\ref{eff2pi6g}(a), 
 and $(1+\delta_R)$ is the radiative correction. 
Since MC simulation does not perfectly describe the experimental 
particle losses, we apply an additional correction of  (10$\pm$5)\%,
$\epsilon_{\rm corr}$, determined from the same data using the
$\epem\to\pipi\piz\eta$ reaction, as described in 
ref.~\cite{3pieta}.  

To calculate the inclusive cross section for the process
$e^+e^-\to \pipi\ppz\eta$, we use events obtained from the 
$\eta$ peak of figure~\ref{chi2}(b), and the  efficiency shown in figure~\ref{eff2pi6g}. 

The radiative correction $(1+\delta_R)$ values for
this process, shown in figure~\ref{eff2pi6g}(b) vs c.m. energy,
are calculated according to ref.~\cite{kur_fad}, iteratively taking into account
the energy dependence of the measured cross section.
The radiative correction has a
very weak energy dependence around $(1+\delta_R)$=0.8 value, and we fit it with the a linear function, and use interpolated values for each experimental point.

The obtained cross section is presented in 
figure~\ref{xs_2pi2pi0eta}(a) and listed in table~\ref{xs_all}.
Our measurements are in  reasonable agreement with the previous
studies performed by the BaBar ~\cite{isr2pi6g} and SND
~\cite{snd2pi6g} experiments.

While the statistical uncertainties in the cross section measurement are relatively large, a structure around the nucleon-anti-nucleon ($N\bar N$) production threshold  is seen.
In the insert we show an energy region around the $N\bar N$ production threshold with a "step" function fit, similar to that in  ref.~\cite{cmdppbar}. The fit yields $\Delta\sigma = 0.15\pm0.05$ nb cross section "step" at the 
$p\bar p$ and $n\bar n$ production thresholds, shown by lines.

Using Eq.~(\ref{xsformular}), we also calculate  the cross section for the 
$\epem\to\omega(782)\piz\eta$ process, shown in figure~\ref{xs_2pi2pi0eta}(b)
in comparison with the previous measurement by BaBar~\cite{isr2pi6g}.
The branching fractions $BR(\omega(782)\to\pipi\piz)=89.2\pm0.07$\%~\cite{pdg} is  taken
into account. Our data for the $\epem\to\omega(782)\piz\eta$ cross
section, listed in  
table~\ref{xs_all}, are in good agreement with the BaBar data. 

\section{Systematic uncertainties and corrections}
\label{syst}
\hspace*{\parindent}

All cross sections above have a 1\% systematic uncertainty from the 
luminosity measurement~\cite{lum}, a 5\%  uncertainty from the
efficiency calculation,  and about 5\% from the MC-data discrepancy in
the efficiency for the charged and  
neutral pions (see Sec.~\ref{eff}), and 1\% from the uncertainty on the 
radiative correction. 
The trigger efficiency, estimated using two  independent trigger
(based on the drift chamber or calorimeter information), is 
 close to unity with an associated systematic uncertainty of 1\%.

The above  uncertainties are combined in quadrature with a 3\% (7\%)  contribution from the 
 variation of signal and background shapes in the fit used to extract
 the $\pipi\ppz\eta$ ($\omega\piz\eta$) signal  yields. The resulting
 total systematic uncertainty for the measured cross 
 sections is 8\% (10\%)

\section*{Conclusion}
\hspace*{\parindent} 
We report the  measurement of the $\epem\to\pipi\ppz\eta$  
cross section with the CMD-3 detector at the VEPP-2000 collider. 
We also present the cross sections for the intermediate state 
$\omega(782)\piz\eta$ which dominates in our energy region. 
Our results are more detailed and in agreement with the previous
experiments performed by the BaBar ~\cite{isr2pi6g} and SND
~\cite{snd2pi6g} collaborations. 

No evidence is  found for the $a_0(980)\to\piz\eta$  resonance in the
$\epem\to\omega\piz\eta$ process within our energy range, in 
 contrast to the BaBar observation at higher center-of-mass energies. However, a   
 small contribution from processes involving $\rho^{\pm}(770)$ resonances has been observed.

\section*{Acknowledgment}
We thank the VEPP-2000 personnel for excellent machine operation. Part
of this work related to the photon reconstruction algorithm in the
electromagnetic calorimeter is supported by the Russian Science
Foundation (project \#14-50-00080).  

\begin{table}[tbh]
\caption{Integrated luminosity, number of signal events, and the $\epem\to\pipi\ppz\eta$ and $\omega(782)\piz\eta$ cross sections vs \Ecm, 
measured with the CMD-3 detector. Only statistical errors are
shown. Lines consecutive separate data from 2020, 2021, and 2022  experimental runs.
}
\label{xs_all}
\vspace{-0.7cm}
\begin{center}
\renewcommand{\arraystretch}{0.85}
\begin{tabular}{cccccc}
\hline
{\Ecm, MeV}
&{Luminosity, nb$^{-1}$}
&{$N(\pipi\ppz\eta)$}
&{$N(\omega\piz\eta)$}
&{$\sigma(\pipi\ppz\eta)$, nb}
&{$\sigma(\omega\piz\eta)$, nb}\\
\hline
1870   & 9342.4  & 161.7$\pm$20.0 & 180.3$\pm$17.6 & 0.83$\pm$0.10 & 1.04$\pm$0.10 \\
1890   & 8955.6  & 174.3$\pm$18.6 & 170.8$\pm$17.7 & 0.92$\pm$0.10 & 1.01$\pm$0.10 \\
1900   & 9721.9  & 204.3$\pm$19.8 & 170.0$\pm$19.3 & 0.99$\pm$0.10 & 0.92$\pm$0.10 \\
1920   & 9941.6  & 261.3$\pm$20.8 & 268.5$\pm$21.9 & 1.22$\pm$0.10 & 1.40$\pm$0.11 \\
1940   & 8908.5  & 209.7$\pm$19.2 & 215.2$\pm$19.8 & 1.08$\pm$0.10 & 1.24$\pm$0.11 \\
\hline
1940   & 5549.1  & 144.1$\pm$27.7 & 117.5$\pm$20.1 & 1.19$\pm$0.23 & 1.09$\pm$0.19 \\
1960   & 11074.4 & 365.9$\pm$31.2 & 307.6$\pm$30.0 & 1.49$\pm$0.13 & 1.40$\pm$0.14 \\
1980   & 10101.1 & 318.9$\pm$34.0 & 297.6$\pm$29.9 & 1.41$\pm$0.15 & 1.47$\pm$0.15 \\
2007   & 21675   & 817.9$\pm$76.0 & 607.4$\pm$41.6 & 1.66$\pm$0.15 & 1.38$\pm$0.09 \\
\hline
1580   & 5836.5  & 0.45 $\pm$7.1  & 5.3  $\pm$6.3  & 0.00$\pm$0.07 & 0.06$\pm$0.07 \\
1600   & 4828.1  & 8.7  $\pm$5.3  & 2.7  $\pm$5.4  & 0.10$\pm$0.06 & 0.04$\pm$0.07 \\
1620   & 4712.8  & 0.7  $\pm$10.0 & 1.0  $\pm$6.2  & 0.01$\pm$0.12 & 0.01$\pm$0.08 \\
1640   & 5923.5  & 1.7  $\pm$7.0  & 0.5  $\pm$5.7  & 0.02$\pm$0.07 & 0.00$\pm$0.06 \\
1660   & 5245.9  & 3.1  $\pm$8.6  & 1.0  $\pm$6.4  & 0.03$\pm$0.09 & 0.01$\pm$0.08 \\
1680   & 19707.2 & 23.8 $\pm$23.1 & 5.1  $\pm$15.6 & 0.07$\pm$0.06 & 0.02$\pm$0.05 \\
1700   & 5256.9  & 14.5 $\pm$9.7  & 5.6  $\pm$8.1  & 0.15$\pm$0.10 & 0.06$\pm$0.09 \\
1720   & 7017.1  & 8.8  $\pm$10.1 & 30.6 $\pm$7.4  & 0.07$\pm$0.08 & 0.26$\pm$0.06 \\
1740   & 5188.9  & 15.1 $\pm$7.8  & 18.5 $\pm$9.1  & 0.15$\pm$0.08 & 0.21$\pm$0.10 \\
1760   & 5014.1  & 49.5 $\pm$10.9 & 29.6 $\pm$10.2 & 0.51$\pm$0.11 & 0.34$\pm$0.12 \\
1780   & 11568.5 & 92.5 $\pm$18.0 & 98.9 $\pm$12.4 & 0.41$\pm$0.08 & 0.49$\pm$0.06 \\
1800   & 5373.7  & 63.8 $\pm$13.6 & 50.2 $\pm$9.7  & 0.60$\pm$0.13 & 0.53$\pm$0.10 \\
1820   & 5834.6  & 66.2 $\pm$11.8 & 55.1 $\pm$11.5 & 0.56$\pm$0.10 & 0.52$\pm$0.11 \\
1840   & 9837.6  & 173.1$\pm$23.1 & 137.3$\pm$15.0 & 0.86$\pm$0.11 & 0.76$\pm$0.08 \\
1860   & 9981.2  & 125.2$\pm$26.4 & 143.5$\pm$18.3 & 0.60$\pm$0.13 & 0.78$\pm$0.10 \\
1870   & 8946.8  & 185.5$\pm$21.9 & 174.2$\pm$21.1 & 0.99$\pm$0.13 & 1.05$\pm$0.13 \\
1872   & 15825   & 303.1$\pm$24.1 & 270.2$\pm$21.1 & 0.92$\pm$0.07 & 0.92$\pm$0.07 \\
1875   & 10615.7 & 159.0$\pm$18.0 & 190.9$\pm$19.1 & 0.72$\pm$0.08 & 0.96$\pm$0.10 \\
1876.6 & 14838.7 & 193.4$\pm$23.2 & 220.5$\pm$22.9 & 0.62$\pm$0.07 & 0.79$\pm$0.08 \\
1877.8 & 10866.6 & 224.7$\pm$23.4 & 224.0$\pm$20.7 & 0.99$\pm$0.10 & 1.10$\pm$0.10 \\
1879.2 & 11122   & 168.3$\pm$18.8 & 213.3$\pm$21.5 & 0.72$\pm$0.08 & 1.02$\pm$0.10 \\
1880.4 & 10743.2 & 175.2$\pm$21.9 & 177.4$\pm$20.3 & 0.78$\pm$0.10 & 0.88$\pm$0.10 \\
1882   & 9790.9  & 192.1$\pm$18.3 & 186.2$\pm$22.4 & 0.93$\pm$0.09 & 1.01$\pm$0.12 \\
1884   & 10094.6 & 139.5$\pm$24.2 & 163.3$\pm$18.8 & 0.66$\pm$0.11 & 0.86$\pm$0.10 \\
1887   & 10078   & 146.4$\pm$23.3 & 157.6$\pm$19.6 & 0.69$\pm$0.11 & 0.83$\pm$0.10 \\
1890   & 12073.1 & 212.0$\pm$22.4 & 204.8$\pm$20.1 & 0.83$\pm$0.09 & 0.90$\pm$0.09 \\
1895   & 10927   & 194.6$\pm$23.0 & 194.7$\pm$17.8 & 0.84$\pm$0.10 & 0.94$\pm$0.09 \\
1897.5 & 7009.4  & 145.0$\pm$18.3 & 163.4$\pm$18.8 & 0.97$\pm$0.12 & 1.23$\pm$0.14 \\
1900   & 5555.6  & 105.5$\pm$15.4 & 113.8$\pm$16.6 & 0.89$\pm$0.13 & 1.08$\pm$0.16 \\
1902   & 5936.3  & 106.2$\pm$13.9 & 110.1$\pm$16.1 & 0.84$\pm$0.11 & 0.98$\pm$0.14 \\
1904   & 5442.9  & 98.6 $\pm$14.9 & 120.9$\pm$16.0 & 0.85$\pm$0.13 & 1.17$\pm$0.15 \\
1906   & 5917.7  & 130.6$\pm$16.7 & 109.2$\pm$16.9 & 1.03$\pm$0.13 & 0.97$\pm$0.15 \\
1908   & 5516.2  & 115.6$\pm$14.4 & 109.9$\pm$13.3 & 0.98$\pm$0.12 & 1.04$\pm$0.13 \\
\hline
\end{tabular}
\end{center}
\end{table}

\end{document}